# Data Analysis of Decision Support for Sustainable Welfare in The Presence of GDP Threshold Effects: A Case Study of Interactive Data Exploration

Fahimeh Asgari[1], Seyedeh Gol Ara Ghoreishi[2], Matin Khajavi[3], Ali Foozoni[3], Ali Ala[4,5], Ahmad Gholizadeh Lonbar[6,*]

*[1] Department of G. Brint Ryan College of Business, University of North Texas, Texas, USA*
*[2]Department of Computer Science, Florida Atlantic University, FL, USA*
*[3]Foster School of Businesses, University of Washington, Washington, USA*
*[4]Department of Mathematics, Saveetha School of Engineering, Saveetha Institute of Medical and Technical Sciences (SIMATS), Tamil Nadu, Chennai, 602105, India*
*[5]School of Mechanical & Materials Engineering, University College Dublin, D04 V1W8, Ireland*
*[6]Department of Civil, Construction, and Environmental Engineering, University of Alabama, Tuscaloosa, AL, USA*
**Corresponding author: Agholizadehlonbar@crimson.ua.edu*

**Abstract**

Energy usage and GDP have been the subject of numerous studies over the past decades. It has been overlooked by previous studies that energy consumption correlates with economic growth in relation to GDP. This study uses threshold regression and Granger causality testing to identify GDP-dependent causality in ten OECD countries over the last 10 years. There is a significant correlation between economic growth and energy consumption. Energy consumption and short-term economic growth are statistically significantly correlated. There is a significant positive effect of energy consumption (EC) on GDP in the short run below the threshold of $10,936 USD since the coefficient is 3.34 and the p-value is 0.0252. There is a -0.0127 correlation coefficient and a 0.0327 p-value associated with GDP Granger-cause EC over the long run. EC and GDP are not causally related for GDP per capita above $10,936 USD. In the long run, GDP Granger causes EC with a coefficient of -0.1638 and a p-value of 0.0675. According to the study, sustainable development requires sustainable use of natural resources, technological investment, foreign direct investment, and gross fixed capital formation. Economic growth can be boosted while adhering to sustainability goals by implementing these recommendations.

**keywords** : Sustainable Welfare, GDP, Interactive Data Exploration model, Energy

## 1- Introduction

Sustainability and well-being are major challenges in modern science and politics. A high standard of living is envisioned for all residents as part of the United Nations'

Sustainable Development Goals without exceeding the limits of the planet. As civilization advances, economic growth has become an increasingly popular method of achieving sustainable development goals that are centered on human welfare. As a result of societal, environmental, and personal factors, the public recognizes the importance of human welfare in human-nature interactions. The viability of the economy and the consumption of energy are essential for long-term growth (Zhang et al. 2024, Hackett et al. 2014). As a result of economic expansion, energy consumption increases, which leads to socioeconomic development, but also causes environmental problems that outweigh the benefits to human welfare (Beckerman et al. 2015, Hussain et al. 2024, Reihanifar. et al. 2023 ). Human welfare is significantly enhanced by energy use. As a result of sustainable development, energy, the environment, and the economy have become critical issues for governments and academia around the world. The concept of human well-being encompasses economic, physical, cultural, and social factors. According to Wang et al. (2013) and Far et al. (2024), energy has a critical role to play in human existence and economic activity, indicating well-being, economic progress, and addressing a fundamental human need. Therefore, a nation's economic growth can be accurately measured by its energy consumption per capita. Energy is seen today as both a manufacturing input and a strategic asset that affects international relations, the global economy, and politics (Pang et al. 2024, Asadi et al. 2024, Reihanifar et al. 2023). Among the most significant economic factors affecting national and international competition are energy supply and procurement conditions. These factors make power one of the most pressing global challenges today. This study discusses the theoretical foundation and research history of this topic to introduce the model and associated tests. As a nation's wealth grows, it can continue to enhance the well-being of its citizens.

## 2- Literature Review

Identifying GDP threshold effects requires interactive data exploration, according to the literature on decision support systems (DSS). Economic stability and welfare outcomes can be improved by data-driven decision-making frameworks, according to numerous studies. Using interactive data exploration, this review synthesizes research findings on the nuanced impact of GDP thresholds on sustainable development, focusing on methodologies and case studies. As a result of forest loss concerns, Heidarlou et al. (2024) investigated forest cover change (FCC) in the Zagros forests (ZFs). An emphasis was placed on the importance of accurate statistics for effective conservation policy-making for the study, which analyzed statistical data on human welfare, biophysical, and climate drivers of FCC in ZFs. The relationship between sustainable human development, gender gaps, welfare, and stock returns between 2012 and 2023 was examined by Korkmaz et al. (2024). Investing in countries with sustainable development and low gender gaps is beneficial for stock returns, according to the study. Sharifi et al. (2024) demonstrated the innovative use of digital twins in engineering urban water systems by applying them to urban drainage systems. Providing a framework for understanding and organizing the existing research landscape while highlighting key areas for future research, this study provides a comprehensive review of terminology, practices, and trends in smart stormwater management. Literature shows that smart technology has benefited stormwater management. Despite advances in quantity management, water quality management remains a challenge. Using digital twins and artificial intelligence in smart city stormwater infrastructure, this study examines the scientific literature on urban drainage systems. Moghim et al. (2023) investigated the role of different cumulus and microphysics schemes in the Weather Research and Forecasting (WRF) model for simulating heavy rainfall and cyclones in Bangladesh.

Ahmadi et al. (2024) integrated a digital twin model with deep learning techniques to create. Nikpour et al. (2024) provided an in-depth review of IoT-based frameworks for managing energy in smart cities, highlighting their capability to track, regulate, and enhance energy usage through advanced analysis. Their study indicates that IoT devices boost efficiency by incorporating real-time monitoring, data processing, and external application integration.  a global terrain and altitude map focusing on Florida's coastline. ROC curve analysis and high AUC values illustrate the model's effectiveness in categorizing terrain precisely. The study's primary goal is to construct a physical and simulation model of Florida's coastline that accurately reflects reality. Additionally to detailed terrain mapping, this approach will benefit environmental monitoring and urban planning. The digital twin model will significantly improve the reliability and performance of geographic information systems. Tran (2022) examined the causal relationship between energy consumption and economic growth in 26 OECD nations. They identified $48,170 USD as the critical value for GDP impact through threshold regression and panel VECM testing. The causal relationship between energy and GDP exists in the first regime, but beyond the threshold, the causal relationship is unidirectional. The Genuine Progress Indicator (GPI) was estimated by Senna et al. (2021) as an alternative to GDP during the period 2002-2016 in Rio de Janeiro. Despite GDP declines from 2014 to 2016, the GPI/GDP ratio remained stable between 22% and 31% using GPI 2.0 methodology. The driving forces behind changes in wetland ecosystem service values (ESVs) in Northeast China between 1980 and 2015 were studied by Song et al. (2021). According to the study, provision and cultural services increased while regulation and supporting services declined using the Emergy method and Logarithmic Mean Divisia Index (LMDI). The impact of energy policy on social well-being was examined by Roach and Meeus (2020). In order to develop a comprehensive strategy for a manufacturing organization's development, they combined energy plans with product development challenges. A quota contribution scheme for tradable green certificates significantly increased clean energy penetration and welfare surplus, according to GAMS optimization applications. Azami and Almasi (2020) examined how energy consumption is related to economic sustainability in oil-producing nations over the long term using the ISEW. According to Granger causality test, energy consumption and economic sustainability have a one-way relationship according to their findings.

Ahmadi (2020) investigated a novel programming-based expert system model in the industry. Reihanifar and Naimi (2016) investigated the Project Value Assessment Engineering (PYDDM) approach, emphasizing its role in optimizing materials, refining work methods, and improving workplace efficiency. Feuerbacher et al. (2018) claim that Bhutan can achieve 100% organic status by using a comprehensive computing general equilibrium model. In spite of Bhutan's low initial reliance on agrochemicals, organic farming would significantly reduce Bhutan's GDP and welfare. A non-separable hybrid DEA model was used by Cai et al. (2019) to analyze China's regional total factor energy efficiency. There was a potential energy savings of 30.6% in China from 2012 to 2016, based on the average total factor energy efficiency of 0.694 from 2012 to 2016. Asadi et al. (2025) conducted a comprehensive review of EAHE research, categorizing studies into analytical, numerical, and exergoeconomic approaches. Their findings highlight the role of EAHEs in reducing energy consumption, minimizing greenhouse gas emissions, and improving cost-effectiveness. Gaspar et al. (2017) compared the GDP and ISEW growth rates over the past five years. The energy-growth nexus hypothesis is supported by their findings of a positive correlation between GDP and energy consumption. A disaster may affect development through channels such as haphazard development, weak institutions, and a lack of social safety nets, according to Mochizuki et al. (2014). Genuine Progress Indicator (GPI) is popular throughout various US states, according to Bagstad et al. (2014). Gholizadeh Lonbar et al. (2025) introduced a hybrid decision-making framework that integrates FMEA and Z-number techniques to enhance financial risk evaluation in media systems. Their research highlights the media mix model as a key strategy for planning advertising campaigns, tackling budget distribution complexities across different advertising platforms.

A comprehensive regional well-being approach emphasizes the importance of both identifying economic benefits and identifying costs. Tugcu (2016) proposes the Index of Sustainable Economic Welfare Growth (ISEW) as a replacement for GDP for Sub-Saharan Africa. According to the researchers, ISEW was a good indicator of feedback, whereas GDP was neutral. Table 1 summarizes the aims, methods, and results of a number of studies examining economic growth, energy consumption, environmental responsibility, and sustainable welfare in different regions and over different periods. To assess the impacts of economic activities on social and environmental well-being, a variety of research methods are used, ranging from energy efficiency analysis to deep learning models and GDP per capita comparisons. Methods used to assess the sustainability of Chinese cities were evaluated by Yin et al. (2011). Cao et al. (2024) proposed an MCNF model to optimize airport security queues by reallocating passengers, significantly improving throughput and reducing wait times for scalable, long-term management. The title of National Model City for Environmental Protection was awarded to 71 cities and five municipalities until 2010, and another 124 cities are actively seeking it. Various assessment tools are discussed in the text, including the ecological footprint method, the economic welfare index, and the human development index. In order to assess the sustainability of cities, eco-efficiency was introduced as an additional tool.

**Table 1:** Summary of Recent Studies on Sustainable Welfare and Economic Indicators

| Author | Year | Aim | Method | Findings |
|---|---|---|---|---|
| Qi et al. | 2025 | Examine the impact of renewable energy technology innovation (RETI) on human development. | Empirical study using data from 58 countries (2006-2020). | RETI enhances HDI through energy structure optimization, efficiency improvement, and poverty alleviation, particularly in high-governance countries |
| Hu er al. | 2025 | 2025 | Analyze ecological welfare in the Dual Carbon goal | Shift to multidimensional analysis and empirical research for better evaluation. |
| Damayanti et al. | 2024 | Examine growth and emissions reductions in ASEAN countries from 1971 to 2017 | Energy efficiency analysis | Energy efficiency mitigated emissions growth caused by GDP and population growth; significant emissions increase in Indonesia and Thailand |
| Govindan et al. | 2024 | Utilizing IoT to increase consumer willingness to return e-waste | Developing a mixed-integer linear programming (MILP) model, a heuristic algorithm based on the Lagrangian relaxation, evaluating performance using GAMS and simulated problems, and validating with Iranian company data | Optimized operational and strategic costs through the use of IoT technology; validated model and algorithm with sensitivity analysis |
| Hoseinbor et al. | 2022 | Investigate the relationship between sustainable economic welfare and energy consumption | Deep convolutional neural network (DCNN) and Granger causality models | Economic welfare and energy consumption are significantly correlated |
| Sun et al. | 2022 | Assess eco-efficiency and social well-being factors in 284 Chinese cities | Network data envelopment analysis (DEA) model | Local government contributions are important in assessing eco-efficiency and social well-being; improved public services and social welfare |
| Wang | 2021 | Evaluate environmentally responsible development | PSR methodology and factor weight ranking technique | Evaluation of environmentally responsible development |
| Sharma et al. | 2020 | Investigate how life expectancy, wellbeing, and eco-footprint affect the Happy Planet Index | Ease of Living Index indicators | Using the Ease of Living Index to measure quality of life instead of GDP to reflect national welfare |
| Mishell et al. | 2020 | Quantify ISEW using Ecuadorian data from 2001 to 2015 and compare with GDP as a development indicator | ISEW quantification | Personal consumption plays a crucial role in enhancing long-term happiness; large gap between sustainable welfare and GDP in Ecuador due to natural resource degradation |
| Bithas et al. | 2018 | Explore the concept of decoupling in reducing resource requirements for GDP production | GDP per capita analysis | Decoupling involves reducing resource requirements for GDP production; suggested using GDP per capita for valid comparisons |

According to Chen (2005), China has grown its GDP at an average annual rate of around 10%, and life expectancy and education have improved significantly since 1978. As well as identifying key institutional infrastructures for equitable income distribution and sustainable growth, the study examined both material attainments and non-material economic welfare. SNI refers to net national income (NNI) adjusted for environmental resource use costs, developed by Gerlagh, Reyer et al. (2002). A general equilibrium model was used to calculate the economic costs associated with sustainable development goals in the Netherlands in 1990 based on SNI variants.

## 3- Methodology

A study of OPEC member countries between 2000 and 2020 used econometric analysis and the Granger test to investigate the relationship between energy consumption and economic growth. Data for this applied and developmental study was gathered using the library method and statistics from the World Bank's documents and official website. The Granger causality test was used to determine whether energy consumption and economic growth are causally related. Statistical tools based on the principle that "cause precedes effect in time" are used to determine causal relationships between time series using Granger causality tests. The Granger criterion states that process X causes process Y when the past values of the time series X(t) provide more information than the past values of the time series Y(t) in predicting the future values of the time series Y(t). In order to detect the Granger causal relationship between two time series X and Y, two linear regressions are used. A threshold effect estimation and a Granger causality test are the two main estimation stages for this study. Figure 1 provides an overview of the study's methodology.

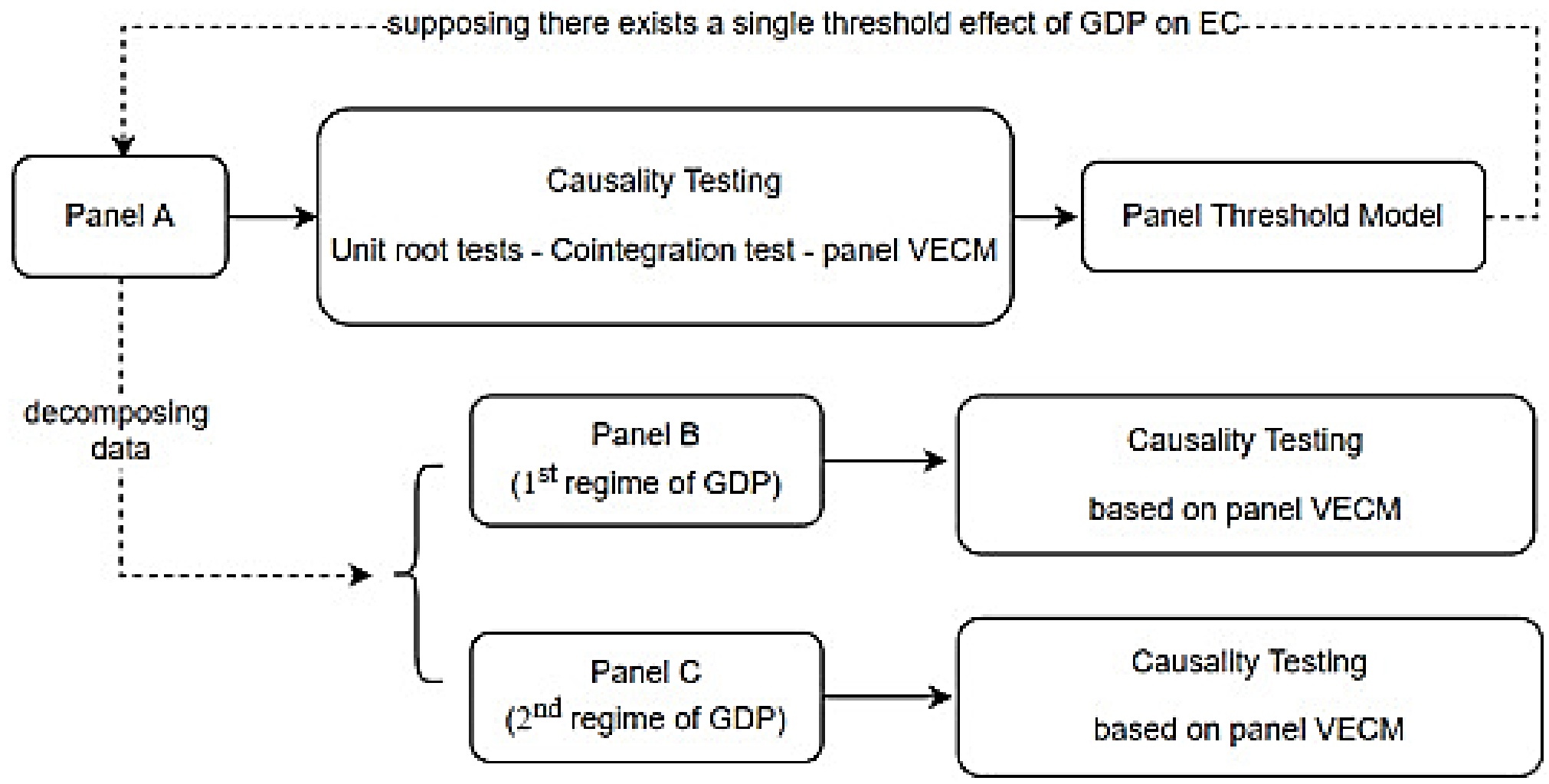

**Fig. 1.** Flowchart of research methodology.

We conduct a causality analysis on the original dataset, followed by three stages of causality testing. The root unit tests are used to analyze the stationarity and integration sequence. Using panel cointegration tests, we can determine whether these variables are cointegrated. The Granger causality between energy consumption and GDP is then assessed using panel VECMs based on fixed-effect or random-effect models. Panel Threshold Model tests for a threshold effect between GDP and energy consumption, searches for two or more regimes endogenously, and estimates of different GDP regimes' effects on energy consumption when economic growth is determined to be a Granger effect. A causality test is repeated for each group of data based on the recommended GDP threshold in order to investigate the GDP-dependent causality.

### *3-1- Panel data*

Panel data can be used to examine issues that cannot be investigated through time series or cross-sectional analysis alone. As a result, cross-sectional data are only useful for identifying economies of scale. Time series data, however, exhibit both effects simultaneously without any separation. It is possible to estimate econometric models and analyze economic trends by combining time series and cross-sectional data. Furthermore, policy decisions can be made based on the results. Panel data offer more information, greater variability, and less collinearity. Time series, however, have a higher collinearity. Because panel data combine time series and cross-sectional data, estimates are more reliable because of the cross-sectional dimension.

## 3-2- Unit root test of panel data

### *3-2-1-Levin, Lin, Chu (IPS) test*

Hypothesis H1 of this test indicates that ρ_i has distinct values. Therefore, the following assumptions are made in order to conduct this test:

$$H_0\{p_i = 0, \qquad i = 1,2,\dots,N \tag{1}$$

$$H_1\begin{cases}\rho_i < 1, & i = 1,2,\dots,N \\ \rho_i = 1, & i = 1,2,\dots,N\end{cases} \qquad 0 < N_1 < N$$

As a result of these assumptions, it is possible for some sections to have a single root. Instead of using combined data for each section, the unit root test is used for each section separately, and the average of these statistics is then calculated as $t_{NT}$, instead of combining the data. The standard statistic $\overline{t_{NT}}$ is defined as follows if $t_{iT}(\pi_i, B_i)$ is the t statistic for the unit root test on the i-th section with an interval of π_i and test coefficients B_i:

$$\overline{t_{NT}} = \frac{1}{N}\sum_{i=1}^{N} t_{iT}(\pi_i, B_i) \tag{2}$$

The value of which tends to infinity and toward the standard normal distribution as N and T increase.

### *3-2-2-Limmer test, Hausman test*

The data we are dealing with may contain both longitudinal and cross-sectional data. In general, such a set of data is referred to as a panel of data or data panel. Panel data models can be estimated in two general situations. In the first case, the width from the origin is the same for all sections, thus a data pool model is used. A data panel mode is characterized by varying widths from the origin for each section. A test called F-Limmer identifies the two above cases. To select between regression methods based on pool data and regression methods based on panel data (combined), the F-Limmer test is used. The following statistics were obtained as a result of this test:

Taking only one width from the origin, the null hypothesis is accepted if the estimated values of F are less than the table value. If the calculated F exceeds the F in the table, the null hypothesis is rejected. Alternatively, group effects are accepted if the calculated F is significantly higher than the table F. Different origins should be taken into account when estimating group effects.

$$F = \frac{\frac{R_{fe}^2 - R_{pool}^2}{n-1}}{\frac{1 - R_{fe}^2}{nt - n - k}} \quad (٣)$$

Regression coefficients R2 depend on the number of logical observations, the width from the common origin, and the time period. Mixed models and panel models are determined by Limer (Chow) tests. If the null hypothesis is rejected, panel data should be analyzed for modeling. It is recommended to use a mixed model otherwise. The Hausman test can be used to determine whether a fixed effect model or a random effect model is appropriate. A fixed effects model is used if the null hypothesis is rejected; otherwise, a random effects model is used.

### *3-2-3- Reliability test of variables*

One of the most significant topics in the analysis of time series data is the discussion of stationarity. Prior to discussing mana, it is necessary to distinguish between strict mana and weak mana. Generally, a strictly Mana process is one in which the probability of the sequence {Yt} is equal to the probability of the sequence {Yt+k} for every k. Weak mean processes are defined as follows: a time series that meets the following three conditions is said to be a weak mean process. The following conditions must be met: the mean, variance, and autocorrelation coefficients must remain constant over time. As a result, this means:

$$E(Y_t) = \mu \quad (4)$$

$$Var(Y_t) = E(Y_t - \mu)^2 = \sigma^2 \quad (5)$$

$$Corr(Y_t, Y_{t-k}) = \frac{\gamma_k}{\sigma^2} = \rho_k \tag{6}$$

To estimate the coefficients of the desired models, traditional and usual econometric methods were used. As a result, variables in the desired model are assumed to be weak, or stable.

### *3-2-4- Dickey-Fuller test (DF)*

The Dickey-Fuller test is one of the methods that are used to perform the unit root test. A first-order self-explanation process is considered in this method:

$$Y_t = \rho Y_{t-1} + \varepsilon_t \quad t = 1, 2, \ldots \tag{7}$$

In the Dickey-Fuller test, the equation above is estimated after subtracting from the sides, resulting in:

$$H_0 : \alpha = 0 \tag{8}$$

$$H_1 : \alpha < 0 \tag{9}$$

Therefore, in the case of the time series instability test, the null hypothesis and the opposite hypothesis for the time series instability test would be as follows:

$$Y_t - Y_{t-1} = (\rho - 1) Y_{t-1} + \varepsilon_t \tag{10}$$

$$\Delta Y_t = \alpha Y_{t-1} + \varepsilon_t \tag{11}$$

In the case of the t statistic, the usual statistic to use is the one that is calculated as follows:

$$t = \frac{\hat{\alpha}}{Se(\hat{\alpha})} \tag{12}$$

There is an estimate of the coefficient as well as a standard error of the estimate. Under the validity of the null hypothesis of the existence of a single root, the statistic (t) presented by (3-14) does not have a normal probability distribution, nor does it have a standard form.

### *3-2-5- Panel cointegration test*

The two-time series xt and yt are considered cumulative with order (d, b) d ≥ b> 0. A linear relationship such as a1yt + a2xt, with the order (d-b), exists between two time series with the accumulation order d. This can be expressed mathematically as follows:

$x_t \sim I(d)$ , $y_t \sim I(d)$

$$\Rightarrow y_t , x_t \sim CI\ (d , b) \qquad (13)$$

$\sim I\ (d\text{-}b) \qquad a_1y_t + a_2x_t$

An acronym for co-accumulation is CI. A cointegration vector is the linear combination of two time series, i.e. [a1, a2]. Co-accumulation is a concept in economics that describes the relationship between two or more time series variables over time based on theoretical foundations in order to form a long-term equilibrium relationship. There is a tendency for these series of data to follow each other randomly over time, but they also follow each other well enough to maintain their differences over time. Co-accumulation, therefore, implies long-term equilibrium between the economy and its environment. In addition to being unstable, panel data also require accumulation and testing. As with stationarity tests, panel data cointegration tests are more powerful than cointegration tests for cross-sectional units separately. Despite their short duration and small sample size, these tests can be applied to a wide range of situations. Cointegration tests for panel data are discussed in this part of the research. Researchers tested the clustering of panel data using Dickey-Fuller statistics after estimating the long-term relationship between variables.

$$DF_\rho = \frac{\sqrt{NT}(\hat{\rho}-1)+3\sqrt{N}}{\sqrt{10.2}}$$

$$DF_t = \sqrt{1.25t_\rho} + \sqrt{1.875N} \qquad (14)$$

ρ is the regression coefficient of the long-term error on the interval of errors arising from the estimation of the model employing the combination approach ($e_{it}$).

$$\hat{e}_{it} = \rho\ \hat{e}_{it-1} + \mu_i \qquad (15)$$

N in the statistics DFt and DFρ represent the number of sections, whereas t is the standard t value of the relationship's coefficient. The retrieved statistics have a normal distribution with a mean of zero and a standard deviation of one. The following assumptions underlie the panel data clustering test:

$$H_1 : \rho < 1 \quad \& \quad H_0 : \rho = 1 \qquad (16)$$

In all cross-sectional units, the zero coefficient shows the absence of co-accumulation across variables, whereas the null hypothesis suggests the presence of co-accumulation.

### *3-2-6- Panel threshold regression model*

1. Theoretical regression model

An analysis of threshold regression models can be used to determine whether economic growth affects energy consumption in a country, according to Hansen et al. A single-panel threshold model has the following structure:

$$y_{it} = \mu_i + \beta_1' x_{it} I(q_{it} \leq \gamma) + \beta_2' x_{it} I(q_{it} > \gamma) + e_{it} \qquad (17)$$

Considering only autocorrelation as an explanatory variable, the k vector of $x_{it}$ is the vector of linear regression explanatory variables for individuals. Equation (4) can be rewritten as follows if the error term is independent and identically distributed:

$$y_{it} = \mu_i + \beta' x_{it}(\gamma) + e_{it} \qquad (18)$$

where $\beta' x_{it}(\gamma) = \begin{cases} \beta_1' x_{it} I(q_{it} \leq \gamma) \\ \beta_2' x_{it} I(q_{it} > \gamma) \end{cases}$

Based on whether it is below or above the threshold value, the sample can be divided into two regimes. Different regression coefficients distinguish the regimes.

$$\overline{y_i} = \mu_i + \beta' \overline{x_i}(\gamma) + \overline{e_i} \qquad (19)$$

where $\overline{y_i} = 1/T \sum_{t=1}^{T} y_{it}, \overline{x_i} = 1/T \sum_{t=1}^{T} x_{it}$, and $\overline{e_i} = 1/T \sum_{t=1}^{T} e_{it}$

The average value of (4.1) has increased over time

$$y_{it}^* = \beta' x_{it}^*(\gamma) + e_{it}^* \qquad (20)$$

where $y_{it}^* = y_{it} - \overline{y_i}, x_{it}^*(\gamma) = x_{it}(\gamma) - \overline{x_i}(\gamma)$, and $e_{it}^* = e_{it} - \overline{e_i}$

Each of its elements is contained within this equation. A vector representation of (21) can be seen in Eq 4:

$$y_{it}^* = \begin{bmatrix} y_{i2}^* \\ \vdots \\ y_{iT}^* \end{bmatrix}, x_{it}^*(\gamma) = \begin{bmatrix} x_{i2}^*(\gamma) \\ \vdots \\ x_{iT}^*(\gamma) \end{bmatrix}, \text{ and } e = \begin{bmatrix} e_{i2}^* \\ \vdots \\ e_{iT}^* \end{bmatrix} \qquad (21)$$

By stacking data and errors over individuals, one time period can be eliminated.

$Y^*, X^*$, and $e^*$, i.e., $y_{it}^* = \begin{bmatrix} y_{i2}^* \\ \vdots \\ y_{iT}^* \end{bmatrix}, x_{it}^*(\gamma) =$ (22)

$\begin{bmatrix} x_{i2}^*(\gamma) \\ \vdots \\ x_{iT}^*(\gamma) \end{bmatrix}$, and $e = \begin{bmatrix} e_{i2}^* \\ \vdots \\ e_{iT}^* \end{bmatrix}$, the model for

## 4- Empirical Regression Model

In order to evaluate the panel threshold model, we assume that economic growth and energy consumption have a single threshold effect:

$$Lnwelfare = \mu_i + \alpha_1 \text{LnGDP } P_{it} I(\text{LnGDP } P_{it} \leq \gamma) + \alpha_2 \text{LnGDP } P_{it} I(\text{Ln } GDP_{it} > \gamma) + \theta_1$$
$$\text{Ln}CPI\ _{it}\ I_{it} + \theta_2 . GGF_{it} + \theta_3 . EC_{it} + \theta_4 . Urban_{it} + \theta_5 . IM_{it} + \theta_6 . EX_{it} + \varepsilon_{it} \quad (23)$$

i and t represent the country and time (year), respectively. Heterogeneity among countries is the fixed effect ($\mu_i$), whereas energy consumption is EC. GDP is the actual gross domestic product per person. The consumer price index, or CPI, reflects energy prices. The proportion of the population living in urban areas compared to the total population is called the $Urban_{it}$ percentage. *A $GGF_{it}$ is a measure of the general government's final consumption expenditures (current LCU).* EX is a variable that represents exports, while IM is a variable that represents imports. In accordance with the threshold equation, energy use is affected by GDP as thresholds are crossed (Table 2).

**Table 2.** Variables included in the descriptive statistics

| Variable | Mean | Std. Dev. | Min. | Max. |
|---|---|---|---|---|
| *Welfare* | 5674.78 | 5656.45 | 456.76 | 845.85 |
| *GDP* | 4673.10 | 4993.44 | 343.45 | 932.43 |
| *EC* | 3148.44 | 2264.66 | 346.11 | 561.12 |
| *CPI* | 7645.68 | 4753.48 | 378.66 | 767.66 |
| *URBAN* | 5066.76 | 5253.18 | 278.78 | 978.83 |
| *GGF* | 4850.88 | 4766.78 | 368.89 | 756.08 |
| *IM* | 6857.86 | 3456.96 | 467.94 | 524.07 |
| *EX* | 5678.96 | 2235.53 | 434.05 | 856.67 |

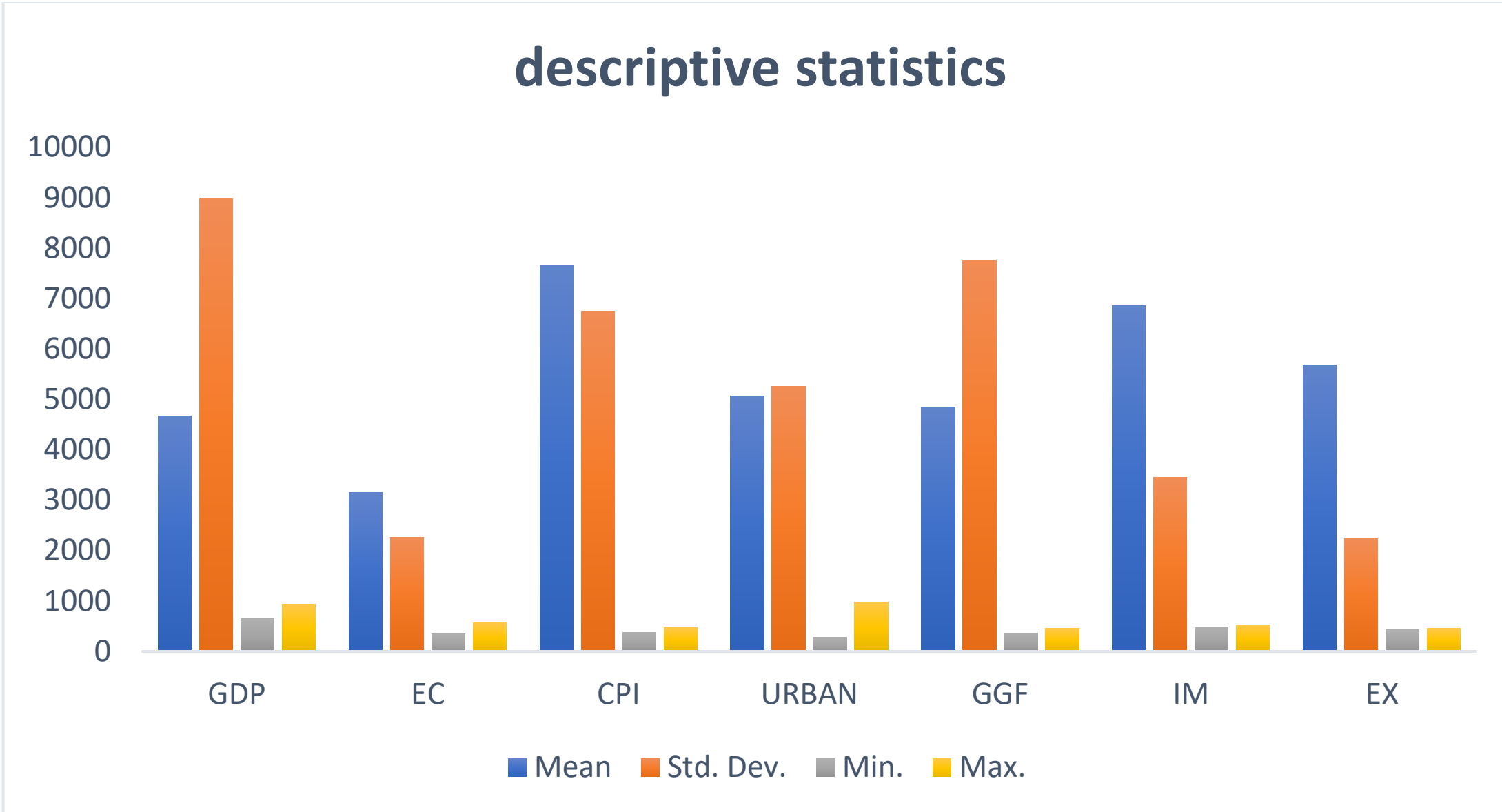

**Fig. 2:** The descriptive statistics results

Tables (1) and Figures (2) present descriptive statistics for the research variables. It is important to consider the mean as the most important index when assessing the centrality of the data since it represents the equilibrium point and center of gravity of the distribution. As another descriptive parameter, the standard deviation can be used to indicate how dispersed the data is. A further point to note is that the minimum and maximum parameters shown in the above table indicate the range of change in the data over time. In this case, the midpoint is the center of the data, which is smaller than half of the data, and larger than the other half, which represents the center of the data. Listed above are the averages, medians, highest values, lowest values, and standard deviations for each of the research variables that were examined during this study. Dispersion is measured by the standard deviation, which is the mean squared distance between each data point and the mean. Since the significance level of all variables is less than 0.05, it can be concluded that the distribution of variables is not normal (Table 3).

**Table3:** The results of the unit root test for research variables (without differentiation - with one-time differentiation)

| Variable name | The value of the statistic | probability value | result |
|---|---|---|---|
| *Ln Welfare* | 5.6798 | 0.0004 | *Confirmation* |
| *Ln GDP* | 3.4586 | 0.0052 | *Confirmation* |
| *Ln EC* | 8.5678 | 0.0030 | *Confirmation* |
| *Ln CPI* | 8.8972 | 0.0001 | *Confirmation* |
| *Ln URBAN* | 9.0857 | 0.0000 | *Confirmation* |
| *Ln GGF* | 7.4550 | 0.0002 | *Confirmation* |
| *Ln IM* | 5.5462 | 0.0003 | *Confirmation* |
| *Ln EX* | 7.6771 | 0.0002 | *Confirmation* |

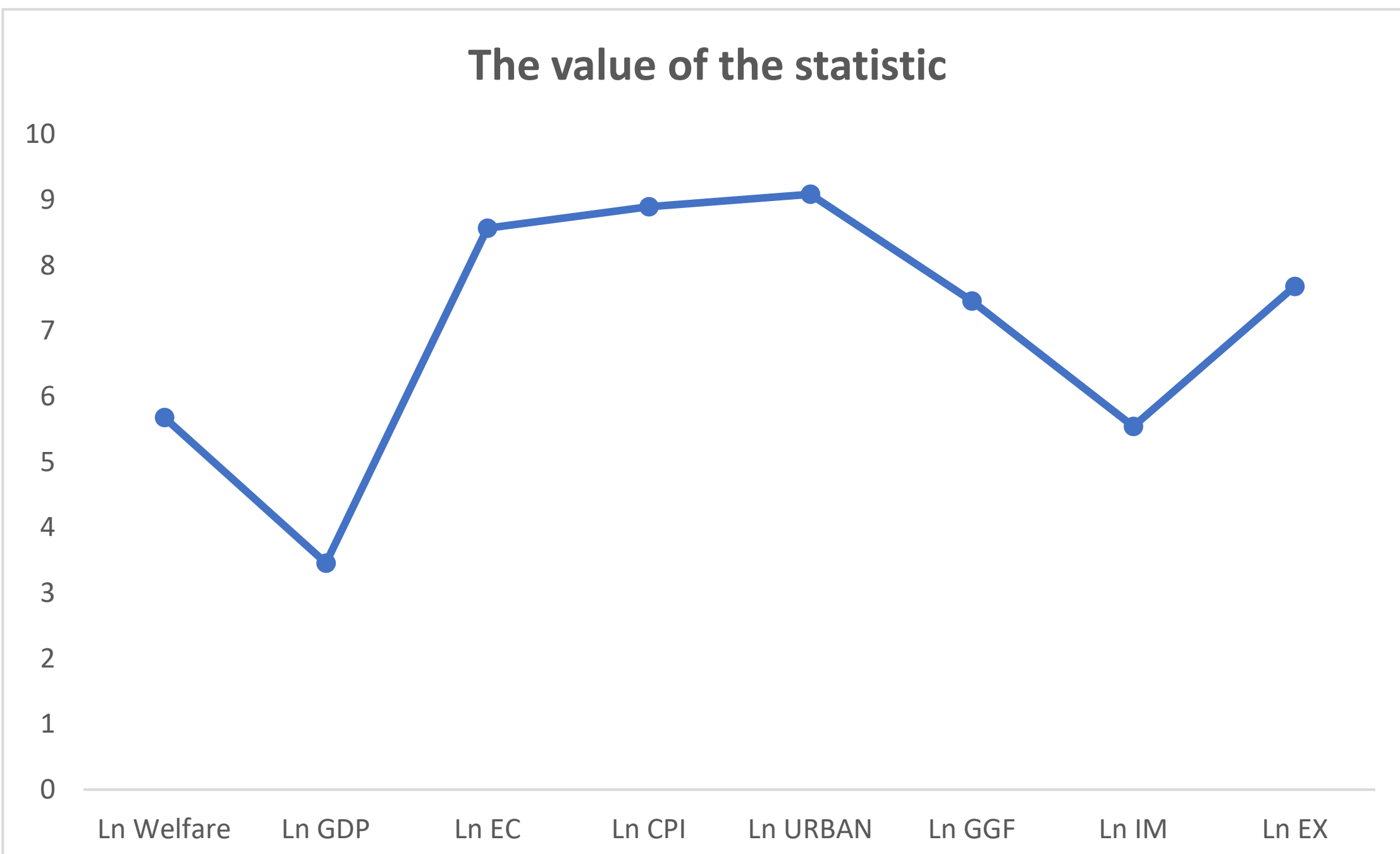

**Fig 3.** The results of the unit root for the value of the statistic

The following are the outcomes of the co-accumulation test of Kao residuals on the variables of the aforementioned model (Fig 3):

**Table 4:** Cointegration test results

| Name of the test | The value of the t statistic | probability value |
|---|---|---|
| Cointegration test of Kao residuals | 5.45632 | 0.0034 |

In each of these models, the Kao co-accumulation test statistic probability value is less than 0.05. As a consequence, the premise that the model variables have no long-term link is rejected (see Table 4).

One of the prerequisites of regression is the lack of collinearity between the explanatory variables in the model, hence this issue is managed by constructing the correlation matrix prior to evaluating the model. To examine the collinearity of the model's explanatory variables, the correlation coefficient and its significance level are computed. The correlation coefficient matrix appears as follows (see Table 5).

**Table 5**: Correlation matrix between explanatory variables

| | *Ln GDP* | *Ln EC* | *Ln CPI* | *Ln URBAN* | *Ln GGF* | *Ln IM* | *Ln EX* |
|---|---|---|---|---|---|---|---|
| *Ln GDP* | 1.00000 | | | | | | |
| | ----- | | | | | | |
| *Ln EC* | 0.05657 | 1.000000 | | | | | |
| | 0.09577 | ----- | | | | | |
| *Ln CPI* | 0.34543 | 0.47306 | 1.000000 | | | | |
| | 0.06568 | 0.00030 | ----- | | | | |
| *Ln URBAN* | 0.14954 | 0.30508 | 0.46767 | 1.000000 | | | |
| | 0.00800 | 0.00046 | 0.00002 | ----- | | | |
| *Ln GGF* | 0.55943 | 0.57070 | 0.56963 | 0.65617 | 1.000000 | | |
| | 0.00480 | 0.00006 | 0.00000 | 0.00006 | ----- | | |
| *Ln IM* | 0.40786 | 0.82078 | 0.75670 | 0.78067 | 0.80068 | 1.000000 | |
| | 0.04906 | 0.00000 | 0.00000 | 0.00008 | 0.00020 | ----- | |
| *Ln EX* | 0.58734 | 0.82020 | 0.35640 | 0.88006 | 0.87007 | 0.67617 | 1.000000 |
| | 0.00557 | 0.00000 | 0.00440 | 0.00020 | 0.00002 | 0.00006 | ----- |

With the exception of the dependent variables, all variables are correlated two-by-two. Correlations are evaluated in degrees, while significant probabilities are quantified in numbers. Since the variables are not strongly related, there is no collinearity between them. Additional numbers between -1 and 1 are assigned to each residence. Positive correlations are implied by higher numbers, while negative correlations are implied by smaller numbers.

### *4-1-F-tests result*

In a null hypothesis test, the test statistic should have an F distribution. To determine which model fits the population (from which the sample was drawn) best, this test is frequently used in statistical comparisons of models fit to data sets. Significant "F-tests" are generated when data are fitted using the least squares approach. Hypothesis H0, which states that origins have the same width, is compared against hypothesis H1, which states that origins have different widths. In each segment, the origins have the same breadth, according to hypothesis H0. The study hypotheses, as well as the ability to merge data and employ the combined regression model, will be statistically validated. Each will be tested (individually or cumulatively). In the event that the H0 hypothesis is denied, however, the research hypotheses will be examined using panel data. The Eviews software executes this test, which checks the breadth of a model from its origin, and yields the following results:

**Table 6:** The result of the F-Limer test

| **Test type** | **The value of the t statistic** | **probability value** | **result** |
|---|---|---|---|
| F-Limer's test | 43.458 | 0.0000 | It is a panel model (with fixed or random effects). |

In the present investigation, the probability value of the Limer test is less than 0.05. The null hypothesis that there are no fixed or random effects in pooled or cumulative regression is consequently rejected. There is no impact pooling, and they are random (Table 6).

***4-2-Hausmann test result***

The Hausman test decides which estimating approach is more appropriate (fixed detection or randomness of cross-sectional unit differences) based on the results of Flemer's tests for each of the hypotheses. Based on hypothesis H0, the model with random effects is preferable (individual effects are not correlated with explanatory variables), while hypothesis H1 indicates that the model with fixed effects is superior (individual effects and explanatory variables are correlated). Since we know the model has effects and is a panel, we must perform this test. After executing the Hausman test (Table 7), the following outcomes will be achieved:

**Table 7:** Hausman test result

| Test type | Chi-square statistic value | probability value | result |
|---|---|---|---|
| Hausman test | 9.86894 | 0.0003 | The model does not have random effects (has fixed effects) |

As a consequence, the Hausman test reveals a probability value of less than 0.05, refuting the statistical assumption that the model contains random effects. Consequently, the regression model has fixed effects on the sections (in this case, the sections of the refinery corporation) and is evaluated using these modifications (Table 8). There is a strong correlation between the welfare variable and GDP.

**Table 8:** Regression model

$$Lnwelfare = \mu_i + \alpha_1 \mathrm{LnGDP}\ P_{it} I(\mathrm{LnGDP}\ P_{it} \leq \gamma) + \alpha_2 \mathrm{LnGDP}\ P_{it} I(\mathrm{Ln}\ GDP_{it} > \gamma) + \theta_1 \mathrm{Ln}CPI\ _{it}\ I_{it} + \theta_2 . GGF_{it} + \theta_3 . EC_{it} + \theta_4 . Urban_{it} + \theta_5 . IM_{it} + \theta_6 . EX_{it} + \varepsilon_{it}$$

| variable | Regression coefficient | Standard error | t statistic | probability value |
|---|---|---|---|---|
| *Ln GDP* | 7.956790 | 2.67980 | 3.124838 | 0.0011 |
| *Ln EC* | 6.075821 | 3.50728 | 9.829868 | 0.0000 |
| *Ln CPI* | 5.560783 | 3.89900 | 1.961075 | 0.0020 |
| *Ln URBAN* | 6.718973 | 4.85146 | 2.080405 | 0.0032 |
| *Ln GGF* | 4.433248 | 3.04797 | 5.455861 | 0.0006 |
| *Ln IM* | 2.665610 | 1.31475 | 2.89947 | 0.0549 |
| *Ln EX* | 2.545856 | 1.78080 | 3.124838 | 0.0593 |
| *Coefficient of determination* | 0.88936 | | | |
| *R-squared* | 0.9669 | | | |
| *Durbin-Watson Test* | 28.65378 | | | |
| *The significance of the model* | 1.89510 | | | |

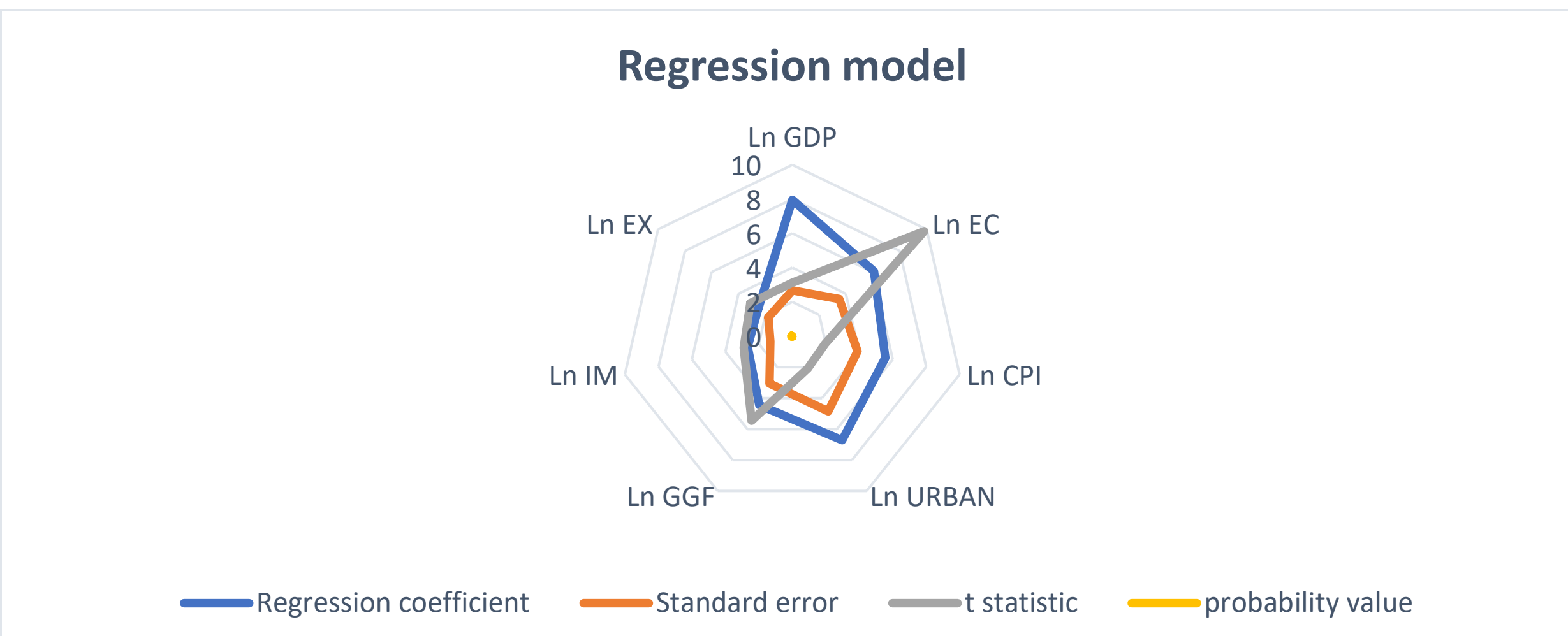

**Fig 4.** The results of the Regression model

GDP increases by 7.956790% when welfare increases by 1%. Recent swings in the flow of products and services have prompted investors to question whether oil businesses still represent a wise investment (Fig 4). based on the given R2 of 0.9669 indicates that the model explains the data very well. An increase in stock prices and attractive dividend income may make investors wealthy during periods of rising oil and gas prices. By acquiring new technology-based equipment, these funds can be used to increase oil and gas production. In the long run, debt repayment, stock buybacks, and dividend payments might contribute to an increase in welfare. For every 1 percent increase in welfare, the energy consumption variable increases by 6.075821 percent (since its probability value is less than 0.05). The price of energy carriers can contribute to the expansion of the welfare index by increasing the energy efficiency of an industrial unit. Based on the p-value, there is a statistically significant association between the consumer price index and the dependent variable. The welfare index should increase by 5.560783 percent with a 1 percent increase in consumer prices. This variable is likewise significant, and consumption uncertainty affects happiness in a significant way. the association between the welfare index and the variables of the Percentage of the total population living (urban) and General government final consumption expenditure is considerable and direct. The Welfare index shows no significant link with import or export factors, as seen in Table 8.

## *4-3- Granger's empirical results*

### *4-3-1- Estimation results for VECM panels by Granger*

The causal link between X and Y will be established once two time series are integrated I(1) and cointegrated. GDP and EC are studied using the panel vector error correction model (VECM). Wald's test was used to test the null hypothesis of short-term causation. To determine the significance of the ECT coefficient, the long-run causality is tested using a t-test. Based on the Granger causality test (Table 9), Table 5 displays the long-run and short-run results.

**Table 9.** Results of the panel VECM Granger causality test.

| **Independent** | | **Dependent** | | **Cause-and-effect understanding** |
|---|---|---|---|---|
| | | ΔEC | ΔGDP | |
| Short-run | ΔEC | – | 2.74(–) (0.3403) | GDP → EC |
| | ΔGDP | 3.87(+)* (0.0032) | – | |
| Long-run | ECTt–1 | –0.0121 [F = 0.43] (0.4532) | –0.0042** (F = 5.21) (0.0322) | EC → GDP |

*, ** indicate significance at 10% and 5%, respectively.

Panel data for ten OECD nations demonstrate a short-run Granger causality between economic expansion and energy consumption. There is evidence that energy use and GDP are causally related. GDP equations with negative coefficients (speed of adjustment) are consistent with a 5 percent long-run equilibrium convergence.

### *4-3-2- Model results for panel thresholds*

Using panel unit root tests, we check whether the variables in their natural logarithmic form are stable at the level in order to estimate Eq. 23.

### *4-3-3- Effects of thresholds on testing*

In Hansen's threshold regression model, the panel threshold model (PTM) is a logarithmic estimator. Table 10 provides the threshold effect test.

**Table 10**. Test results for threshold effect

| Single threshold test | |
|---|---|
| F1 | 137.61 |
| P-value | 0.001 |
| Critical factors (10 %, 5 %, 1 %) | 232.56, 231.02, 215.23 |
| Double threshold test | |
| F2 | 236.54 |
| P-value | 0.053 |
| Critical factors (10 %, 5 %, 1 %) | 132.21, 142.02, 622.63 |
| Test of triple thresholds | |
| F3 | 54.81 |
| P-value | 0.352 |
| Critical factors (10 %, 5 %, 1 %) | 162.24, 115.14, 234.01 |

Economic growth and energy use exhibit a short-run positive unidirectional Granger causation based on panel data for 10 OECD nations. Evidence suggests that energy usage and GDP are causally linked. GDP equation with a negative coefficient (speed of adjustment) is consistent with the long-run equilibrium convergence of 5 percent.

### *3.5.1. Proposed GDP regimes and causal relationships between growth and energy use*

Following the determination of the GDP threshold, we examine GDP regime-dependent causality. According to the suggested GDP criteria, the data may be classified into two distinct categories (GDP per capita 10,936 and GDP 10,936). A Granger causality test is then conducted on each data set using panel VECM. Table 11 compares two proposed GDP regimes. The relationship between energy consumption, GDP, and EC is unidirectional when GDP per capita is less than or equal to $10,936 USD. As the per capita GDP reaches $10,936, long-term and short-term causality separate. There is no causal relationship between the two variables in the near term. There is, however, a unidirectional causal relationship between GDP and energy consumption in the long run. Based on these estimates, we infer that energy consumption and GDP are causally related to GDP regimes, and we suggest distinct economic strategies based on different phases of economic development.

**Table 11**. Results of panel VECM for GDP-dependent causality.

| GDP per capita | Independent variable | | Dependent variable | | Sense of the causality |
|---|---|---|---|---|---|
| | | | ΔEC | ΔGDP | |
| GDP ≤ 10,936 | Short-run | ΔEC | | 3.34 (–)∗∗ (0.0252) | EC → GDP |
| | | ΔGDP | 1.13 (+) (0.3658) | | |
| | Long-run | ECTt–1 | 0.0273 [F = 1.25] (0.2746) | –0.0127∗∗ [F = 5.11] (0.0327) | EC → GDP |
| GDP> 10,936 | Short-run | ΔEC | | 1.05 (–) (0.3064) | EC ↮ GDP |
| | | ΔGDP | 1.87 (+) (0.2137) | | |
| | Long-run | ECTt–1 | –0.1638∗ [F = 4.67] (0.0675) | –0.0095 [F = 1.76] (0.1852) | GDP → EC |

∗ and ∗∗ denote significance at 10% and 5%, respectively.

Economic growth is stimulated when real GDP per capita is below a threshold (i.e., 10,936 USD). In the course of a nation's economic development, energy conservation initiatives may have both short- and long-term effects on GDP. As energy consumption negatively impacts real GDP per capita on a short-term basis, an increase in energy use negatively impacts real GDP per capita on a long-term basis. Statistic evidence indicates a lack of short-run causality in the second GDP regime, but substantial long-term causality over the long run. It will, however, become more pronounced as time passes. As a result, policymakers must consider long-term consequences when making decisions. When real GDP per capita crosses a given threshold, it doesn't change. The final row of Table 11 shows a long-run unidirectional causal relationship between real GDP per capita and energy consumption. The implementation of energy conservation rules may not hinder economic growth.

## 5. Discussion

Global economics and politics are heavily influenced by energy, which is a major input and strategic commodity. It also affects nations' production structures and their competitiveness on national and international markets. Economic growth is affected by several factors, including the budget, the current account deficit, welfare expenditures, and the rate of economic growth. In the current period, the logarithm of energy consumption is influenced by the logarithms of GDP and GDP from the previous period, while prices affect the quantity of energy consumption. Energy consumption increases as energy prices fall. As a result of low energy prices and outdated technology, the current period's energy consumption is impacted by the previous period's energy consumption, which also affects the gross domestic product. Even though energy consumption has a positive impact on economic growth, inflation has a negative effect on it. Economic growth is hindered by the inflation rate because of its influence on it. In accordance with the threshold effect, levels of energy consumption are correlated with levels of gross domestic product. Discrepancies in energy consumption between industrialized and developing nations, as well as between different degrees of economic growth within a country, are predicted to be influenced by economic development.

## 6. Conclusion

Numerous studies have examined the relationship between energy use and GDP over the last few decades. Previous studies have not examined the causal relationship between energy use and economic growth. The threshold regression technique and Granger causality testing using VECMs were applied to a panel dataset of 10 OECD countries from 2000 to 2020. To avoid impeding economic development, energy conservation measures must take economic development into account. Through a few examples, this study examines global energy-saving techniques. Using panel data to determine regime-dependent causality is not possible, according to our understanding. We use threshold regression and Granger causality testing to solve the problem. Causality tests may be affected by the unequal or insufficient nature of separated data sets. In order to understand how GDP threshold affects energy-growth causation, additional research should be conducted. Due to energy conservation and climate change, environmental degradation and resource waste have increased in recent decades. Developing nations need to be more resource efficient. Future research should focus on energy efficiency and energy structure in order to understand the causal relationship between GDP and energy. According to the study, OPEC member states have also benefitted from the energy consumption shock. Enhanced energy efficiency and reduced energy consumption are the results of the feedback effect, which conserves energy. Researchers have found that the rebound effect does not necessarily result in a decrease in happiness. Buyers take advantage of price dispersion in markets for things they highly value or can get relatively cheaply when inflation rates are high. If these conditions are not met, inflation can severely reduce welfare.

### Conflict of interest

In this paper, the authors did not receive funding from any institution or company and declared that they do not have any conflict of interest.